\documentclass{ws-ijmpa}
\usepackage[super,compress]{cite}
\usepackage{graphicx}
\usepackage{xspace}
\usepackage{listings}
\usepackage{url}
\usepackage{natbib}
\usepackage{hyperref}
\usepackage[utf8]{inputenc}

\begin{document}
\markboth{G. Principe}{The time-domain gamma-ray sky seen by the \textit{Fermi}-LAT}

\catchline{}{}{}{}{}
%
%%%%%%%%%%%%%%%%%%%%%%%%%%%%%%%%%%%%%%%%%%%%%%%%%%%%%%%%%%%%%%%%%%%%

\title{The time-domain gamma-ray sky seen by the \textit{Fermi}-LAT}

\author{Giacomo Principe
}

%\from{ins:x}\from{ins:y}\from{ins:z}, \textit{on behalf of the Fermi-LAT Collaboration}}
%\address{Dipartimento di Fisica, Universit\'a di Trieste, I-34127 Trieste, Italy, \\ Istituto Nazionale di Fisica Nucleare, Sezione di Trieste, I-34127 Trieste, Italy, \\
% INAF - Istituto di Radioastronomia, I-40129 Bologna, Italy.\\
% On behalf of the \textit{Fermi}-LAT collaboration.\\ 
% giacomo.principe@ts.infn.it}
 
\address{Dipartimento di Fisica, Universit\`a di Trieste, I-34127 Trieste, Italy \\
Istituto Nazionale di Fisica Nucleare, Sezione di Trieste, I-34127 Trieste, Italy \\
INAF -- Istituto di Radioastronomia, I-40129 Bologna, Italy \\
\vspace{0.3cm}
On behalf of the \textit{Fermi}-LAT Collaboration \\
\texttt{giacomo.principe@ts.infn.it}}

\maketitle

\begin{history}
\received{Day Month Year}
\revised{Day Month Year}
\end{history}

\begin{abstract}
The \textit{Fermi} Gamma-ray Space Telescope is currently celebrating its 15th anniversary of operation. Since its launch, the \textit{Fermi}-Large Area Telescope (LAT), the main instrument onboard the \textit{Fermi} satellite, has remarkably unveiled the sky at GeV energies providing outstanding results in time-domain gamma-ray astrophysics. In particular, LAT has observed some of the most powerful transient phenomena in the Universe (such as gamma-ray bursts, blazar flares, magnetar flares, …) enabling the possibility to test our current understanding of the laws of physics in extreme conditions.
In this paper I will review some of the main recent results with a focus on the transient phenomena seen by LAT with a multi-wavelength and multi-messenger connection. 

\keywords{gamma-rays; time-domain.}
\end{abstract}

\ccode{PACS numbers:}

%\tableofcontents

\section{Introduction}
The Large Area Telescope (LAT) is the main instrument onboard the \textit{Fermi} satellite \cite{2009ApJ...697.1071A}. 
The LAT is a gamma-ray detector which collects photons by conversion into electron-positron pairs and has an operational energy range from 20\,MeV to more than 300\,GeV. The instrument comprises a high-resolution converter tracker (for direction measurement of the incident $\gamma$-rays), a CsI(Tl) crystal calorimeter (for energy measurement), and an anti-coincidence shield detector to identify and veto the background of charged particles.
Since its launch in June 2008, the LAT has been surveying the high-energy gamma-ray sky with no performance concerns and a remarkable 98.7\% uptime for Science mission \cite{2021ApJS..256...12A}.

Thanks to the large number of gamma-ray photons collected, almost 2 billion LAT events publicly available at the FSSC\footnote{\textit{Fermi} Science Support Center \url{https://fermi.gsfc.nasa.gov/ssc/}}, several thousands of gamma-ray sources have been detected so far. In particular, in the latest general catalog (4FGL-DR4), which is based on a comprehensive analyses of LAT data collected in the energy range between 50 MeV and 1 TeV during the first 14 years of the mission, more than 7000 gamma-ray sources have been reported \cite{2020ApJS..247...33A,2022ApJS..260...53A}. Similarly, about 200 and 1500 sources have been detected focusing on the LAT low-energy range (30-100 MeV) and high-energy range (10-1000 GeV), respectively \cite{2018A&A...618A..22P,2017ApJS..232...18A}.

\section{Transient sources seen by \textit{Fermi}-LAT}

\subsection{Active Galactic Nuclei}
Two primary inquiries in active galactic nuclei (AGN) research revolve around the formation of jets and the source of gamma-ray emissions. A favoured approach to explore their origins involves coordinated multiwavelength (MWL) campaigns, which have the capability to discern the intricate details of these source. This methodology can be employed in galaxies that have undergone significant evolution, allowing angular resolution down to the core size (and potentially even to the event horizon, as demonstrated in studies like \cite{2019ApJ...875L...1E,2024A&A...681A..79E}), or in galaxies with newly formed jets (of the order of approximately 100-1000 years old), facilitating the examination of the initial phases of jet formation \cite{2020A&A...635A.185P,2021MNRAS.507.4564P}. Another area of interest, concerning the investigation of AGN variability, involves probing the origins of astrophysical neutrinos \cite{2018Sci...361.1378I}.

A recent and remarkable study is represented by the EHT-MWL observational campaigns on the radio galaxy M87 performed in the last years. During the observations carried on April 2017, the source was found to be in a historically low state. Although it was not possible to disentangle the region responsible for the gamma-ray emission, assuming leptonic processes it was possible to clearly exclude that the EHT region is responsible  for that \cite{2021ApJ...911L..11E}. On contrary, in April 2018 the source underwent a gamma-ray flaring episode, although in the preliminary studies the exact location remains obscure, the observed variability allowed to constrain the size of the gamma-ray emitting region \cite{2024A&A...692A.140A}.

\begin{figure*}[ht!]
\begin{center}
\includegraphics[width=0.98\textwidth]{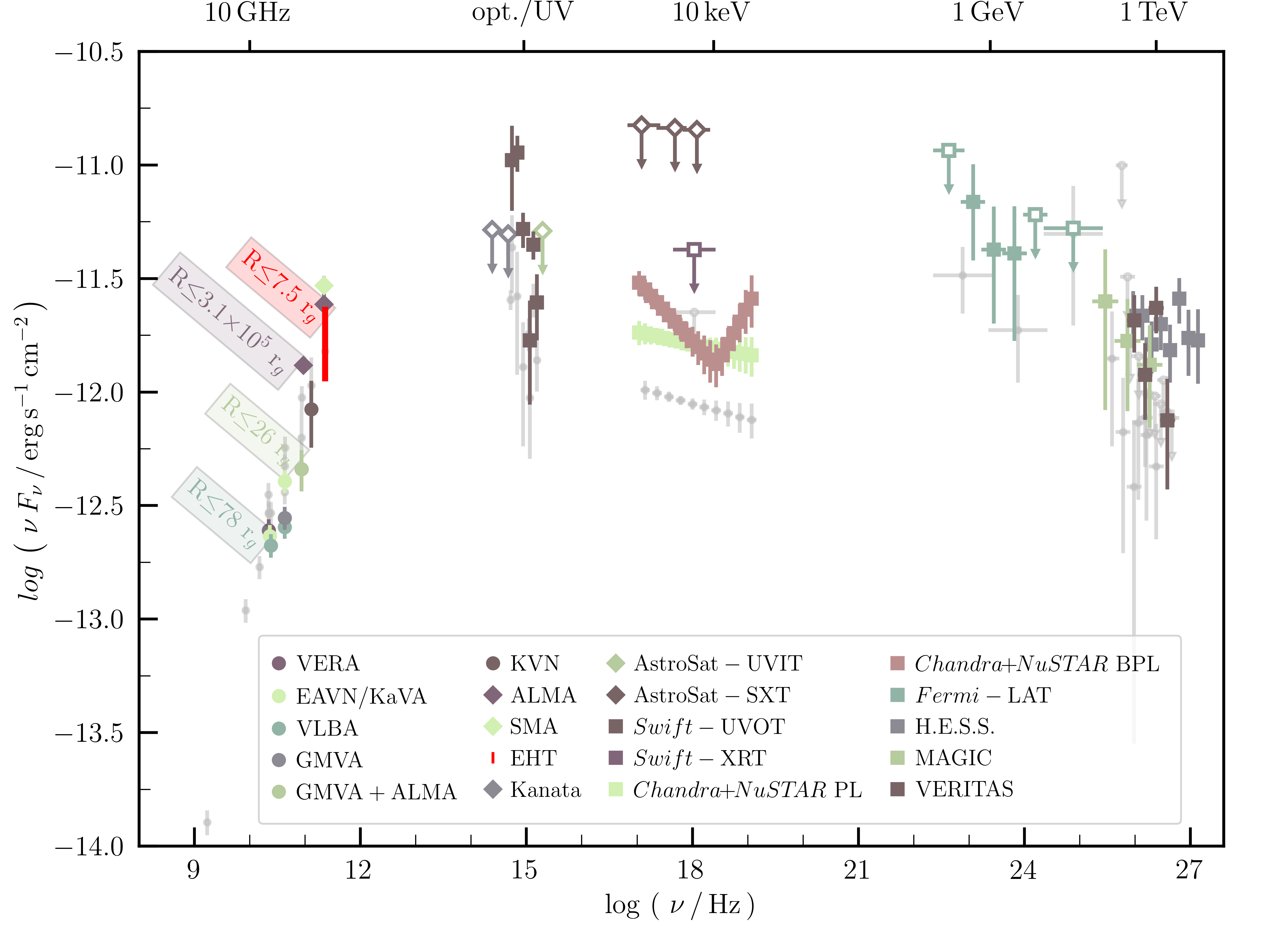}
    \caption{Observed broadband SED of M87, contemporaneous with the EHT campaign in April 2018, with fluxes measured by various instruments highlighted with different colours and markers. Filled markers represent flux point estimates, while empty markers indicate ULs. In the X-ray band, the two sets of points represent the core flux from the models with a power law (green) and broken power law (purple). Grey points represent the observed broadband SED of the 2017 EHT campaign \cite{2021ApJ...911L..11E}. MAGIC observations were performed before the $\gamma$-ray flare, VERITAS observations only partially overlap the flare, while all H.E.S.S. observations have been taken during the VHE $\gamma$-ray flaring episode. For further information see \cite{2024arXiv240417623T}. }
\label{fig:sed}
\end{center}
\end{figure*}

Fig. \ref{fig:sed} shows the 2018 MWL broadband SED as well as the 2017 EHT-MWL campaign SED - displayed in grey in the background - for comparison.

%Neutrino follow-up searches
Moreover, AGNs are among the possible sources for the astrophysical neutrinos detected on Earth. The association of a neutrino with a flaring blazar TXS\,0506+056 \cite{2018Sci...361.1378I} has sparked interest in identifying additional counterparts.
So far, no other gamma-ray counterpart has been unambiguously identified.
However one source raised particular interest as a promising neutrino emitter: PKS\,1502+106; a bright FSRQ located at redshift of z$ = $1.84 \cite{2021ApJ...912...54R}.
A continuous monitoring of the gamma-ray sky has been performed for a few years, with \textit{Fermi}-LAT playing a key role in the identification of candidate counterparts to realtime neutrino alerts. In this study, through a full population investigation using of gamma-ray blazars detected by \textit{Fermi}-LAT, the relationship between the neutrino and gamma-ray luminosities was also evaluated, finding different trends between the two blazar classes BL Lacs and flat-spectrum radio quasars \cite{2024arXiv240106666G}.

\subsection{Gamma-Ray Bursts}
One of the largest class of transient sources detected by the LAT is represented by  gamma-ray bursts (GRBs).
The latest (second) catalog of GRBs detected by the \textit{Fermi}-LAT collaboration was recently published, covering the initial 10 years of its operations from August 4, 2008, to August 4, 2018 \cite{2019ApJ...878...52A}. This catalog comprises 186 GRBs, with 91 exhibiting emissions within the 30$-$100 MeV range (17 of which are exclusively observed within this range) and 169 being detected above 100 MeV. Among the identified bursts, 17 have been categorized as short GRBs, while 169 are classified as long GRBs.

An extraordinary observation was achieved on October 9 2022, marked by the discovery of GRB 221009A, the highest flux GRB ever observed by the \textit{Fermi} Gamma-ray Burst Monitor (\textit{Fermi}-GBM) \cite{2023ApJ...952L..42L}.
This special GRB was also detected by LAT at GeV energies \cite{2022GCN.32637....1B} and it will be the subject of a forthcoming publication as well as included in a proceeding in this journal.

\subsection{Magnetar flares, fast radio bursts, supernovae}
Magnetars are strongly magnetized (magnetic fields of $10^{13-15}$ G) neutron stars rotating on short periods of 0.1-10 s \cite[see e.g.;][]{2015RPPh...78k6901T,2017ARA&A..55..261K}.
While they show few outbursts (flare and pulsating tail) in X-rays and soft gamma-rays, no GeV observations was observed until recently. On April 15th 2020, a magnetar located in the nearby Sculptor Galaxy underwent a giant-flaring episode \cite{2021Natur.589..207R,2021Natur.589..211S} and for the first time GeV-energy emission was detected by \textit{Fermi}-LAT for this class of sources \cite{2021NatAs...5..385F}.

\begin{figure*}[ht!]
\begin{center}
\includegraphics[width=0.98\textwidth]{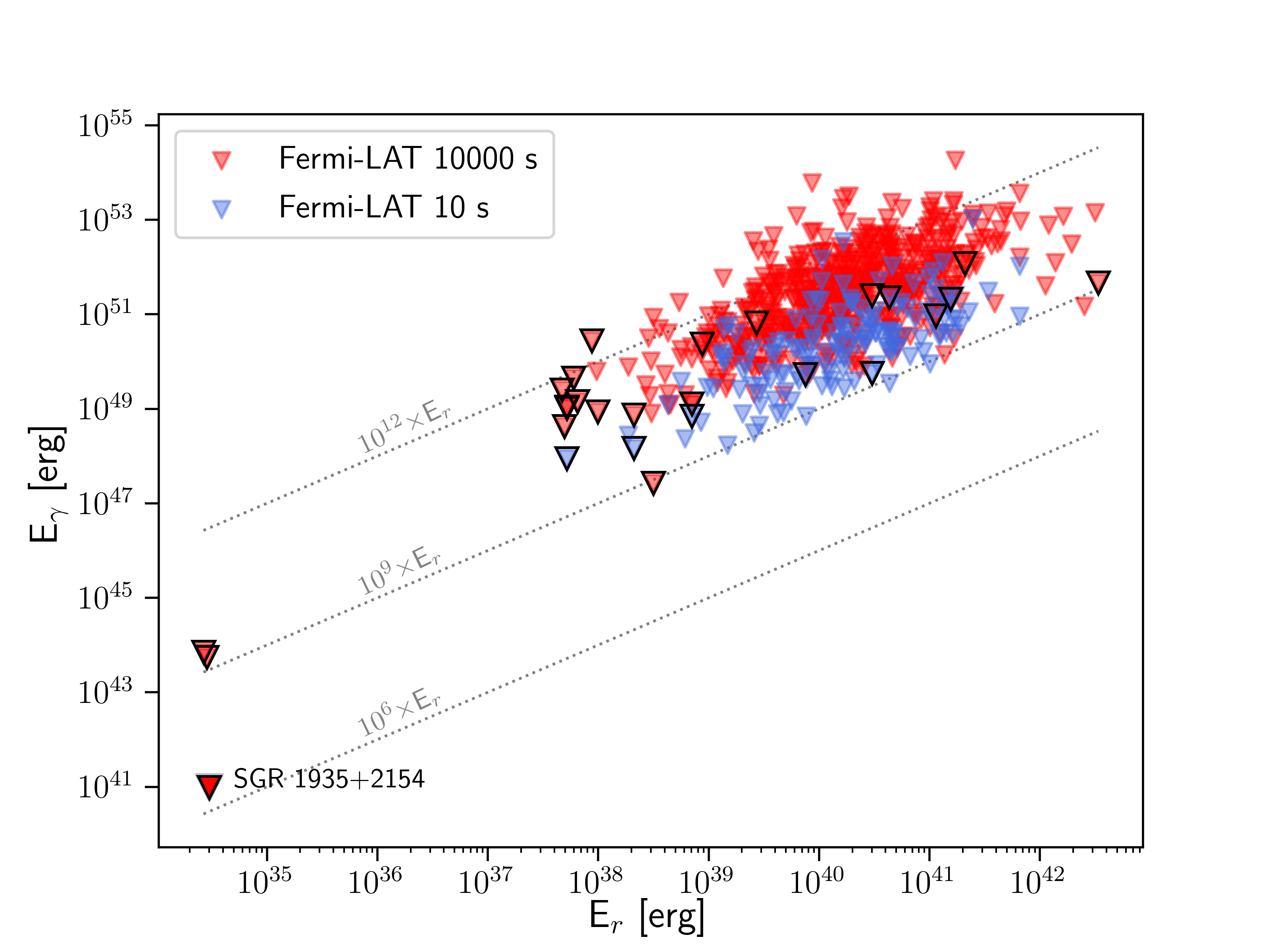}
    \caption{Diagram of the upper limits on the gamma-ray energy flux as a function of radio energy fluxes. The FRB events reported are those with available information on their radio flux density and with a correctly estimated gamma-ray flux UL (i.e. sufficient gamma-ray exposure). Gamma-ray upper limits from the analysis on 10 s (10 000 s) are plotted in blue (red). FRB sources with known host-galaxy are highlighted with a black outline. For further information see \cite{2023A&A...675A..99P}. }
\label{fig:frb}
\end{center}
\end{figure*}

Being fast radio bursts (FRBs) possibly connected with magnetars \cite{2020Natur.587...54C}, searches for gamma-ray emission from repeating and non-repeating FRBs were recently carried out using 12 years of \textit{Fermi}-LAT data \cite{2023A&A...675A..99P}.
No gamma-ray emission has been observed to date, with the last study representing the largest and deepest systematic search for gamma-ray emission from FRBs at gamma-ray energies performed so far (see Fig. \ref{fig:frb}).

For years, supernova remnants (SNRs) have been regarded as the primary sources of Galactic cosmic-ray (CR) acceleration. However, core-collapse supernovae (CCSNe) might also play a significant role in accelerating particles in the initial phases of their development, potentially impacting the overall energy distribution of cosmic rays within the Galaxy.
Recently, the observation of SN2023ixf \cite{2023TNSAN.119....1P}, a type II CCSNe in the nearby galaxy M101, provided the possibility to investigate the CR acceleration in supernovae.
While no significant gamma-ray emission from SN2023ixf was found, based on reasonable assumptions, it was possible to constrain the maximum efficiency of CR acceleration to be as low as 1\%, which contradicts the commonly assumed value of 10\% for typical supernovae\cite{2024arXiv240410487M}.

\subsection{Gamma-ray pulsars}
Before the launch of the \textit{Fermi} satellite in 2008, only a handful of pulsars, including the Crab, Vela, and Geminga, were known to emit gamma-rays \cite{1999ApJS..123...79H}. In the latest catalog released by the LAT collaboration, almost 300 pulsars are observed to emit gamma-rays at GeV energies \cite{2023ApJ...958..191S}. Within this group, 50\% of the gamma-ray pulsars are young. Among these, the subset that remains unobserved in radio transmissions displays a wider distribution across Galactic latitudes compared to their youthful counterparts emitting radio signals. The remaining pulsars are millisecond pulsars (MSPs), six of which are not detected in radio frequencies.

\begin{figure*}[ht!]
\begin{center}
\includegraphics[width=0.98\textwidth]{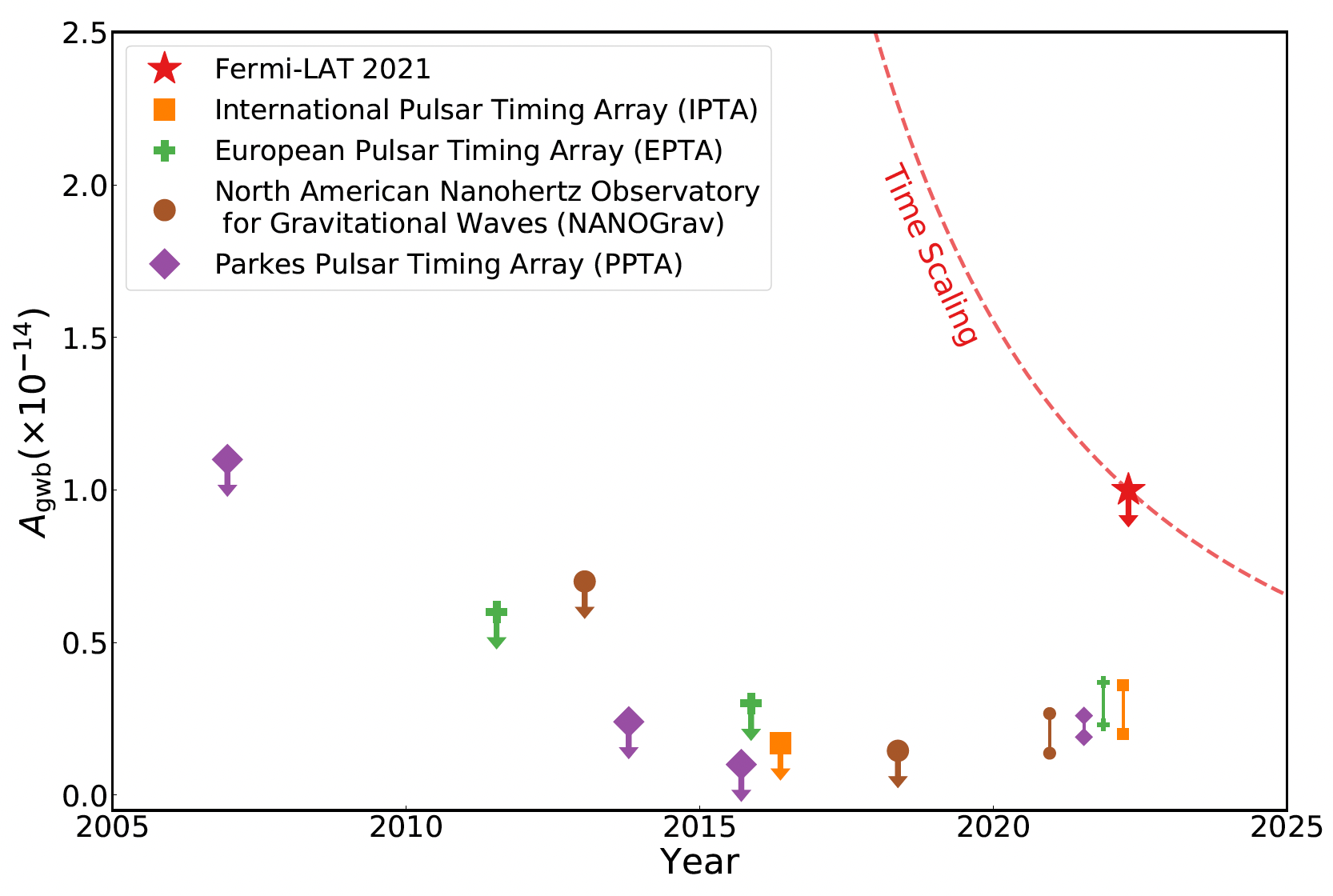}
    \caption{Constraints on the gravitational wave background from radio and gamma-ray
PTAs. The \textit{Fermi}-LAT
95\% upper limit, 1.0$\times$10$^{-14}$, uses data obtained through January, 2021 and is shown as a red
star at an approximate publication date April, 2022. The dashed red line indicates the expected
scaling as the limit as a function of time. The plot is taken from \cite{2022Sci...376..521F}. }
\label{fig:fermi_gw}
\end{center}
\end{figure*}

Relative to the pulsar science, gravitational waves within the extremely low frequency range of 3-100 nHz, generated by the merging of supermassive black hole (SMBH) binaries, were longly predicted to be detectable via pulsar timing array. Specifically, this detection involves monitoring the arrival times of regular pulses emitted by individual pulsars, which may deviate from their expected timing due to disturbances in space-time. In June 2023, for the first time, the detection of a gravitational wave background (GWB) signal was reported with radio pulsars timing array (PTA) \cite{2023ApJ...951L...6R,2023ApJ...951L...8A},

Similarly, a first study at gamma-ray energies was performed using 12.5 years of LAT data to establish a gamma-ray PTA comprising 35 prominent gamma-ray pulsars. This analysis aimed to constrain the GWB \cite{2022Sci...376..521F}. 
Notably, while this pioneering investigation yielded only an upper limit, with accumulation of more data this could provide independent detection on the characteristic strain of the GWB. Fig. \ref{fig:fermi_gw} illustrates the constraints on the GWB derived from gamma-ray data acquired by \textit{Fermi}-LAT, compared to past radio-based PTAs limits.

\subsection{Solar flares}

The nearest celestial source emitting gamma rays is represented by our Sun. The first solar gamma-ray flare was observed on June 11, 1991 by the EGRET instrument \cite{1993A&AS...97..349K}. 
In 2021, the LAT collaboration unveiled the first catalog of solar flares identified by the \textit{Fermi}-LAT \cite{2021ApJS..252...13A}. This compilation encompasses 45 solar flares detected within the energy range of 30 MeV to 10 GeV during the period spanning 2010 to 2018, corresponding to Solar Cycle 24. Notably, this significantly amplifies by almost ten times the set of solar flares detected above 30 MeV. Noteworthy inclusions are the primary identification of GeV emissions emanating from solar flares originating in active regions situated beyond the visible edge of the Sun. While electron emissions dominate conventional solar flares, those detected by LAT exhibit characteristics aligning with a curved model, such as a power-law with an exponential cutoff or a pion decay model, suggesting the involvement of proton and ion acceleration processes.

\section{Conclusion}
\textit{Fermi}-LAT is working without major problems ($>95\%$ uptime for Science) and continues to deliver exciting science results. After more than 15 years of data taking, the focus moved to the search and discovery of (new) transient phenomena as well as the connection to multi-messenger astronomy. 
Specifically, in 2020, a milestone was achieved with the detection of a giant flare from a magnetar at GeV energies.
Relative to this topic, despite the most extensive systematic search for gamma-ray emissions from FRBs, no significant detections were made, but rather stringent upper limits were obtained, which serve to rigorously constrain their origins and emission mechanisms.
Accompanied to the first detection of the GWB using radio PTAs, recently, a study detailing the search for gravitational waves using the \textit{Fermi} gamma-ray pulsar timing array was published. No significant emission was detected but with more statistic it might be possible to detect them in the next decade.
Remarkably, on October 9, 2022, we recorded the brightest GRB to date (an accompanying proceeding in this journal will be dedicated on this event).
While Fermi findings continue to be indispensable for ongoing endeavours in the search for multi-messenger counterparts, they also lay the groundwork for future gamma-ray observations.

\section*{Acknowledgments}
We acknowledge ISCRA for awarding this project access to the LEONARDO supercomputer, owned by the EuroHPC Joint Undertaking, hosted by CINECA (Italy). G.P. acknowledge support by ICSC – Centro Nazionale di Ricerca in High Performance Computing, Big Data and Quantum Computing, funded by European Union – NextGenerationEU. 
%The \textit{Fermi} LAT Collaboration acknowledges generous ongoing support from a number of agencies and institutes that have supported both the development and the operation of the LAT as well as scientific data analysis. These include the National Aeronautics and Space Administration and the Department of Energy in the United States, the Commissariat à l'Energie Atomique and the Centre National de la Recherche Scientifique / Institut National de Physique Nucléaire et de Physique des Particules in France, the Agenzia Spaziale Italiana and the Istituto Nazionale di Fisica Nucleare in Italy, the Ministry of Education, Culture, Sports, Science and Technology (MEXT), High Energy Accelerator Research Organization (KEK) and Japan Aerospace Exploration Agency (JAXA) in Japan, and the K. A. Wallenberg Foundation, the Swedish Research Council and the Swedish National Space Board in Sweden.

Additional support for science analysis during the operations phase is gratefully acknowledged from the Istituto Nazionale di Astrofisica in Italy and the Centre National d'Etudes Spatiales in France. This work is performed in part under DOE Contract DE-AC02-76SF00515.

%\begin{thebibliography}{000} %for 3 digits
%\begin{thebibliography}{00}  %for 2 digits
%\begin{thebibliography}{00}  
%%journal paper
%\bibitem{jpap} R. Loren and D. B. Benson, {\it J. Comput.System Sci.} {\bf 27}, 400 (1983).
%\bibitem[Principe et al.(2018)]{2018A&A...618A..22P} Principe, G., Malyshev, D., Ballet, J., et al.\ 2018, \aap, 618, A22. doi:10.1051/0004-6361/201833116

%\end{thebibliography}

%\bibliographystyle{ws-ijmpa}
\bibliographystyle{plainnat}
\bibliography{bibliography}

\end{document}